\documentclass[twocolumn, amsmath,amssymb,amssymb,pra, showpacs]{revtex4-1}
\usepackage{graphicx,epsfig,epstopdf}
\usepackage{mathrsfs}
\usepackage{dcolumn}
\usepackage{bm}
\usepackage{braket}
\usepackage{color}
\usepackage{xr}
\begin{document}
\title{Time-and-frequency gated photon coincidence counting; a novel multidimensional
spectroscopy  tool}
\author{Konstantin E. Dorfman}
\email{Email: kdorfman@uci.edu}
\author{Shaul Mukamel}
\email{Email: smukamel@uci.edu}
\affiliation{University of California, Irvine, California 92697-2025}
\date{\today}

\begin{abstract}
Coherent multidimensional
optical spectroscopy techniques are  broadly applied  across the electromagnetic spectrum ranging from NMR  to the UV. These reveal properties of matter through correlation plots of signal  fields generated in response to sequences of short pulses with variable delays. Here we discuss a new class of multidimensional techniques obtained by time-and-frequency resolved photon coincidence counting measurements of  $N$ photons  which constitutes a $2N$ dimensional spectrum. A compact description of  these signals is developed based on  time ordered superoperators  rather than the normally ordered ordinary operators used in Glauber's photon counting formalism.The independent control of the time and frequency gate parameters reveals details of matter dynamics not available otherwise. Application to an  anharmonic oscillator model with fluctuating energy and anharmonicity demonstrates the power of these signals.
\end{abstract}

\maketitle


\section{Introduction}

Multidimensional spectroscopy measures correlations of matter dynamics during various time intervals controlled by sequences of short pulses to get material information \cite{ern87,muk00,muk15}. These can be distances between spins in NMR, vibrational motions of protein in the infrared and energy migration in photosynthesis in visible regime. Such correlation plots carry qualitatively higher level of information than single interval one-dimensional (1D) techniques. Here we demonstrate how similar ideas  may be extended to single-photon coincidence counting measurements. An $N$ time-and-frequency gated photon measurement provides a $2N$ dimensional parameter $\omega_jt_j$ space. These techniques performed on bulk ensembles or at the single molecule level offer novel windows into molecular events and relaxation processes that are complementary to coherent multidimensional techniques \footnote{The remaining part of the introduction section is based on our earlier work in Ref. \cite{dor12}.}.


A semiclassical formalism for describing the photon counting process was first derived by Mandel \cite{Man58,Man59}. The full quantum mechanical description of field and photon detection was developed by Glauber \cite{Glau07}. A theory of the electromagnetic field measurement through photoionization and the resulting photoelectron counting has been developed by Kelley and Kleiner \cite{kel64}. An ideal photon detector is a device that measures the radiation field intensity at a single point in space. The detector size should be much smaller than spatial variations of the field. 

The response of an ideal time-domain photon detector is independent on the radiation frequency. The joint frequency and time resolution is limited by the Fourier uncertainty $\Delta\omega\Delta t>1$. A naive calculation of such signals without time and frequency gating can work for slowly varying spectrally broad optical signals where we are far from the Fourier limit $\Delta\omega\Delta t=1$ but otherwise may yield unphysical negative signals \cite{Ebe77}. The mixed time-frequency representation for the coherent optical measurements with interferometric or autocorrelation detection can be calculated in terms of a mixed material response functions and a Wigner distribution for the incoming pulses, the detected field and the gating device\cite{Muk96}. Multidimensional gated fluorescence signals for single-molecule spectroscopy have been calculated \cite{Muk11}.

Glauber's theory of photon counting and correlation measurements \cite{Glauber07,Scully97,Mol70} focuses on theradiation field and is formulated in the field space alone; matter is not considered explicitly. Signals are related to  multi-point normally ordered field correlation functions, convoluted with time and frequency gating spectrograms of the corresponding detectors. This approach takes the detected field as given and does not address the matter information and how this field has been generated. Temporally and spectrally resolved measurements can reveal important matter information. An adequate microscopic description where joint matter and field information could be retrieved by a proper description of the detection process is required for e.g  single photon spectroscopy of single molecules  \cite{Fle00,Let10,Rez12}. 

This paper extends the diagrammatic approach developed in \cite{dor12} for calculating time-and-frequency gated  photon counting measurements  shown in Fig. 1. In particular we apply the formalism of \cite{dor12} to calculate $N$-th order photon correlation measurements generated by a fluctuating oscillator model  that represents a generic molecular system with energy and electron transport. The connection with real molecules makes time-and-frequency gated photon counting a practical spectroscopic tool for studying material systems by analyzing correlated optical signals. These observed signals are represented by a convolution of a bare signal that assumes unrestricted  time and frequency resolution and a detection spectrogram that represents the time and frequency gate functions. The detection process is described in the joint field and matter space by a sum over pathways each involving a pair of quantum modes with different time orderings. The signal is recast using time ordered superoperator products of matter and field. In contrast to Glauber theory which is based on normally ordered field operator, the approach of \cite{dor12} employs time-ordered superoperators.  Time ordering in spectral measurements of atomic source fields was studied in detail by Knoll et al. and Cresser \cite{kno86,cre87}. Passive (filter-like) systems in quantum optics has been studied by \cite{kno87}. 
Experimental applications to normal and time ordered intensity correlation measurements have been reported in the seminal work of Kimble et al. \cite{kim77}. According to these studies free-field operators, in general, do not commute with source-quantity operators. This is the origin of the fact that the normal and time ordering of the measured field correlations, according to the Kelley-Kleiner theory \cite{kel64}, are transformed into normal and time ordered source quantities occurring inside the integral representations of the filtered source-field operators.

The paper is organized as follows. To make the technical presentation more comprehensive we first summarize the approach of ref. \cite{dor12}. In Section \ref{sec:S1} we present closed expressions  suitable for spectroscopy applications for time-and-frequency resolved  single photon counting based on matter correlation functions. In Section \ref{sec:S2} we extend the formalism to multiple detections. In Section \ref{sec:tsj} we apply this to calculate photon counting signals from a simple anharmonic oscillator model coupled to a dynamical two-state-jump model and simulate these results in Section \ref{sec:Sim}. We conclude by discussing the relation between  our approach with the commonly used physical spectrum \cite{Ebe77} in Sec. \ref{sec:conclusion}.

\section{Gated single photon counting}\label{sec:S1}

\begin{figure*}[t]
\begin{center}
\includegraphics[trim=0cm 0cm 0cm 0cm,angle=0, width=0.85\textwidth]{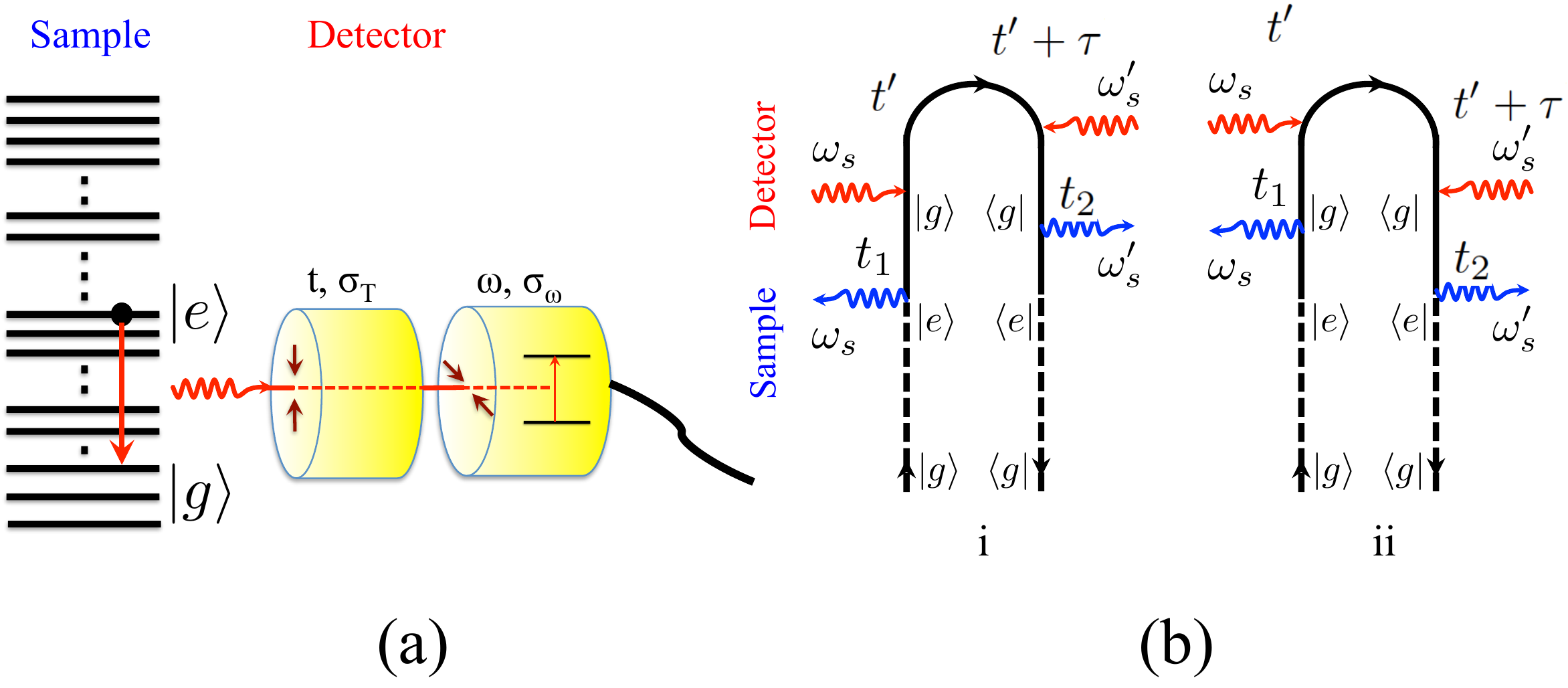}
\end{center}
\caption{(Color online) (a) Schematic of time-and-frequency resolved photon counting signal $S^{(1)}(t,\omega)$. (b) Corresponding loop diagrams that represent the relevant field matter interactions. Red lines correspond to interaction with detector, blue lines represent interactions with molecule. Dashed lines represent arbitrary dynamics of the system prior to the photon emission. Diagram rules are presented in \cite{Rah10}.}
\label{fig:diags1}
\end{figure*}

\begin{figure*}[t]
\begin{center}
\includegraphics[trim=0cm 0cm 0cm 0cm,angle=0, width=0.85\textwidth]{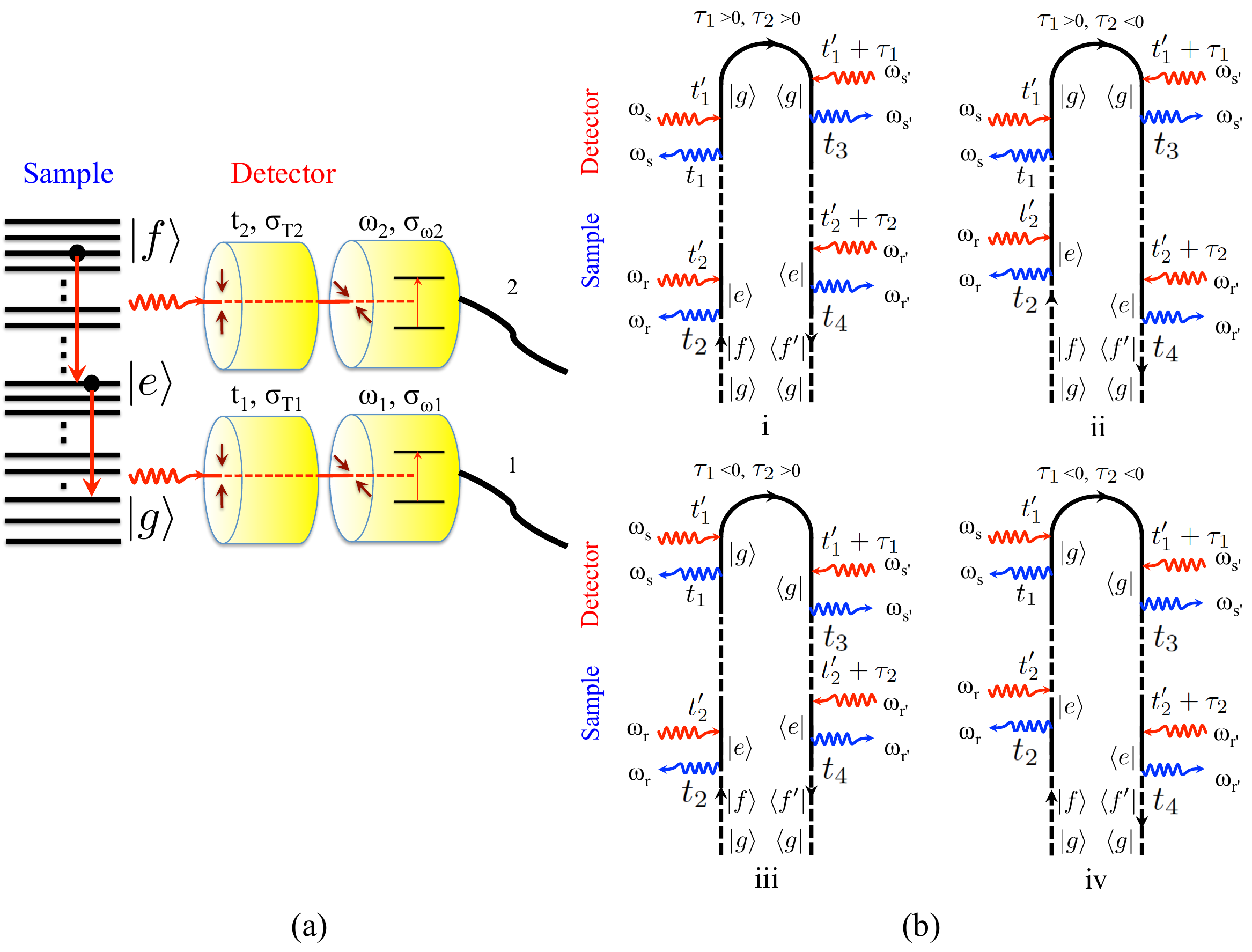}
\end{center}
\caption{(Color online) (a) Schematic of time-and-frequency resolved photon coincidence signal $S^{(2)}(t_1,\omega_1;t_2,\omega_2)$. (b) Corresponding loop diagrams.}
\label{fig:diags2}
\end{figure*}

For spectroscopic applications one has to formulate the signal in terms of matter, rather than field quantities. To connect photon counting signals to the matter response one can use microscopic theory based on the perturbative series over field-matter interactions. Utilizing the superoperator algebra outlined in our earlier papers and its relation with gated photon counting signals (see Fig. 1 a and Appendix \ref{app:field}) one can obtain the gated single photon counting signal (see Appendix \ref{app:S1}) by reading the signal off the diagrams in Fig. 1b
\begin{align}\label{eq:S1g}
S^{(1)}(t,\omega)&=\mathcal{D}^2(\omega)\notag\\
&\times\int dt' \int d\tau D(t,\omega;t',\tau)V^{(1)}(t',\tau),
\end{align}
where $D(t,\omega,t',\tau)$ is a detector time-domain spectrogram  which takes into account the detector parameters given by Eq. (\ref{eq:Ddef}). The detected signal  Eq. (\ref{eq:S1g}) is given by a convolution of the spectrograms of the detector and matter correlation function $V^{(1)}(t',\tau)=\langle \mathcal{T}V_R^{\dagger}(t'+\tau)V_L(t')\rangle$. The detector spectrogram is an ordinary function of the gating parameters whereas the matter correlation function is a superoperator quantity (see Appendix \ref{app:S1})  related to the state of the field and matter prior to detection.

Eq. (\ref{eq:S1g}) can be alternatively recast using time/frequency spectrograms
\begin{align}\label{eq:S1001}
S^{(1)}(t,\omega)&=\mathcal{D}^2(\omega)\notag\\
&\times\int dt' \int \frac{d\omega'}{2\pi} W_D(t,\omega;t',\omega')W_V^{(1)}(t',\omega'),
\end{align}
Here
\begin{align}
W_D(t,\omega,t',\omega')=\int d\tau D(t,\omega,t',\tau)e^{i\omega'\tau}
\end{align}
is a detector Wigner spectrogram and 
\begin{align}
W_V^{(1)}(t',\omega')= \int d\tau e^{-i\omega'\tau}V^{(1)}(t',\tau).
\end{align}
 is the Wigner spectrogram for the first order matter correlation function.

To understand how to construct photon counting signal from matter correlation functions we use the loop diagram shown in Fig. \ref{fig:diags1}b. Two red arrows represent interactions with the detector, whereas two blue arrows correspond to emission from the molecule. The two branches of the loop correspond to ket (left branch) and bra (right branch). During some arbitrary field matter interactions  depicted by dashed lines, the molecule is promoted to an excited state governed by a population density matrix, whereas the field density matrix remains diagonal in the vacuum state $|0\rangle\langle 0|$. After photon emission depicted by two interactions with bra and ket the material density matrix moves to the lower state and the field density matrix is in the state $|1\rangle\langle 1|$. The spectral modes $s$ and $s'$ via their frequency difference $\omega_s-\omega_s'$ are responsible for the temporal bandwidth of the bare signal. If only a single mode photon is absorbed by the detector, all temporal resolution is lost. Similarly, the time difference $\tau$ between the interaction times with the detector yields spectral bandwidth of the photon. Both spectral and temporal bandwidths of the bare photon are then combined with the corresponding bandwidths of the gates and yield the final measurement. These diagrams provide a natural physical picture of the generation of the photon by molecule and its subsequent detection.

\section{Gated multiple photon counting}\label{sec:S2}

 A more elaborate measurement is time-and-frequency resolved photon coincidence counting (PCC) given by Eq. (\ref{eq:S20}). The corresponding diagrams are shown in  Fig. \ref{fig:diags2} a and b, respectively. 
  The gated coincidence signal can be similarly calculated (see Appendix \ref{app:S2})
\begin{align}\label{eq:S2g}
&S^{(2)}(t_1,\omega_1;t_2,\omega_2)=\mathcal{D}^2(\omega_1)\mathcal{D}^2(\omega_2)\notag\\
&\times\int dt_1' \int d\tau_1 D^{(1)}(t_1\omega_1;t_1',\tau_1)\notag\\
&\times\int dt_2' \int d\tau_2 D^{(2)}(t_2,\omega_2;t_2',\tau_2)\notag\\
&\times V^{(2)}(t_1',\tau_1;t_2',\tau_2).
\end{align}
Here $D^{(j)}$ $j=1,2$ indicate the two detector's spectrograms and $V^{(2)}(t_1',\tau_1;t_2',\tau_2)=\langle \mathcal{T}V_R^{\dagger}(t_2'+\tau_2)V_R^{\dagger}(t_1'+\tau_1)V_L(t_1')V_L(t_2')\rangle$ is the relevant second order matter correlation function. The same arguments about temporal and spectral bandwidth discussed about for a single photon detection apply here as well. The corresponding signal recasted using Wigner spectrogram reads
\begin{align}\label{eq:S2001}
&S^{(2)}(t_1,\omega_1;t_2,\omega_2)=\mathcal{D}^2(\omega_1)\mathcal{D}^2(\omega_2)\notag\\
&\times\int dt_1' \int d\frac{\omega_1'}{2\pi} W_D^{(1)}(t_1\omega_1;t_1',\omega_1')\notag\\
&\times\int dt_2' \int d\frac{\omega_2'}{2\pi} W_D^{(2)}(t_2,\omega_2;t_2',\omega_2')\notag\\
&\times W_V^{(2)}(t_1',\omega_1';t_2',\omega_2'),
\end{align}
where
\begin{align}
W_V^{(2)}(t_1',\omega_1';t_2',\omega_2')&=\int d\tau_1e^{-i\omega_1'\tau_1}\int d\tau_2 e^{-i\omega_2'\tau_2}\notag\\
&\times V^{(2)}(t_1',\tau_1;t_2',\tau_2)
\end{align}
is the Wigner spectrogram of the second order matter correlation function.

As can be seen from Eqs. (\ref{eq:S2g}) and (\ref{eq:S2001}) these constitute multidimensional measurements analogous of the $\chi^{(3)}$ nonlinearities detected in rephasing and nonrephasing contributions to the photon echo experiments (see Section \ref{sec:tsj})\cite{Muk95}. Here $\omega_1$ and $\omega_2$ are analogous to Fourier components of the $t$ and $\tau$ delays - $\omega_t$ and $\omega_\tau$, respectively, whereas the time difference between two detectors $t_1-t_2>0$ is analogue of time $t_2$.

The fundamental material quantity that yields the emission spectra (\ref{eq:S1g}) is a two-point dipole correlation function in Eq. (\ref{eq:nVV}) and for the coincidence $g^{(2)}$ measurement (\ref{eq:S2g}) it is four-point dipole correlation function in Eq. (\ref{eq:nnVVVV}). Note that in the perturbative expansion, eq. (\ref{eq:rhoT}), we used superoperator time ordering in both field and matter correlation functions since we expressed the signal using  correlation functions of the field or matter alone.

The normalized $N$- th order photon correlation measurement that describes time-and-frequency measurement performed at $N$ detectors characterized by central time $t_j$ and central frequency $\omega_j$, $j=1,...,N$ is similarly defined as
\begin{align}\label{eq:gN}
&g^{(N)}(t_1,\omega_1,\Gamma_1;...,t_N;\omega_N,\Gamma_N)=
\frac{\langle\hat{n}_{t_1,\omega_1}...\hat{n}_{t_N,\omega_N}\rangle}{\langle \hat{n}_{t_1,\omega_1}\rangle...\langle \hat{n}_{t_N,\omega_N}\rangle},
\end{align}
where $\Gamma_j$, $j=1,...N$ represents other parameters of the detectors such as bandwidth ($\sigma_T^j$ and $\sigma_\omega^j$ are the time gate, and frequency gate bandwidths, respectively). In the absence of gating Eq. (\ref{eq:gN}) reduces to Eq. (\ref{eq:gN0}) . The only difference is the use of superoperator time-ordering rather than normal ordering. However, Eq.(20) directly addresses the photon generation and emission process by the source whereas Eq.(\ref{eq:gN0}) assumes that the field is given.


\section{Application to an oscillator with fluctuating frequency and anharmonicity}\label{sec:tsj}

As an illustration we now apply our  formalism to a toy model of the single anharmonic vibrational mode described by the Hamiltonian
\begin{align}
\hat{H}=\hbar\Omega\hat{B}^{\dagger}\hat{B}+\frac{1}{2}\hbar\Delta\hat{B}^{\dagger}\hat{B}^{\dagger}\hat{B}\hat{B},
\end{align}
where $\hat{B}^{\dagger}$ ($\hat{B}$) are boson creation (annihilation) operators satisfying the commutation relation $[\hat{B},\hat{B}^{\dagger}]=1$. Both frequency $\Omega$ and anharmonicity $\Delta$ are subjected to fluctuations described by coupling to a stochastic bath undergoing a two-state jump TSJ process \cite{Kub63,Sanda1,Konstantin2}
\begin{align}
&\Omega=\Omega_0+\Omega_1\sigma_z,\notag\\
&\Delta=\Delta_0+\Delta_1\sigma_z.
\end{align}
$\Omega_0$ and $\Delta_0$ are the average values, whereas $\Omega_1$ and $\Delta_1$ describe stochastic frequency modulation by chemical exchange represented by two-state-jump (TSJ) model (state up ``$u$'' ($\sigma_z=1$) and down ``$d$'' ($\sigma_z=-1$) where $\sigma_z$ is the Pauli spin matrix.

Three vibrational levels are accessible by third order signals: the ground state $|g\rangle$, the first excited state $|e\rangle=\hat{B}^{\dagger}|g\rangle$ and the doubly excited $|f\rangle=\frac{1}{\sqrt{2}}\hat{B}^{\dagger}|e\rangle$. Their energies are $0$, $\Omega$ and $\Omega+\Delta$, respectively. The corresponding density matrix has nine components denoted $|\nu\nu'\rangle\rangle=|\nu\rangle\langle\nu'|$; $\nu$, $\nu'=g,e,f$. The dipole moment matrix elements are $\mu_{eg}=\mu$ and $\mu_{ef}=\sqrt{2}\mu$.

We shall use the stochastic Liouville equation (SLE) \cite{Kub62,Kub63,Gam95,Tanimura1,Sanda1}  for  modeling spectral lineshapes. It assumes that the observed  quantum system  is coupled to a classical bath that undergoes stochastic dynamics; the bath affects the system but the system does not affect the bath. The bath dynamics is described by a Markovian master equation
\begin{align}\label{eq:ME}
\frac{d\rho}{dt}=\hat{\mathcal{L}}\rho(t)=-\frac{i}{\hbar}[H,\rho(t)]+\hat{L}\rho(t).
\end{align}
The SLE ignores system/bath entanglement but  provides a very convenient level of modeling for lineshapes.  We choose the lorentzian gates (see Appendix \ref{app:gate}). Details of the TSJ model and corresponding signals calculation are presented in Appendix \ref{app:TSJ}.

We start with the $S^{(1)}$ signal  Eq. (\ref{eq:S1g}). The bare signal given by a two-point correlation function of dipole operator $V$  has now to be replaced by a bosonic operators $B$. The two contributions of the signals are depicted as two loop diagram in Fig. \ref{fig:diags1} which are complex conjugate of each other. The signal is given by
 \begin{align}\label{eq:S12}
S^{(1)}(t,\omega)&=2\mathcal{R}\mathcal{D}^2(\omega)(-\hbar^2)|\mu_{eg}|^2\int dt' \int d\tau D(t,\omega;t',\tau)\notag\\
&\times\langle\langle I|\mathcal{G}_{eg,eg}(\tau)\mathcal{G}_{ee,ee}(t')|\rho_{ee} \rangle\rangle_S.
\end{align}
where $\mathcal{R}$ denotes the real part, the initial state in the spin space is spin up state: $|\rho_{ee}\rangle\rangle_S=\rho_{ee}(0)|ee\rangle\rangle\begin{pmatrix}
  1 \\
  0 \\
 \end{pmatrix}$, and we trace over the final state $\langle\langle I|=(1,1)\text{Tr}$, where $\text{Tr}=\langle\langle ff|+\langle\langle ee|+\langle\langle gg|$. Matrix multiplication for the matter correlation (\ref{eq:S2mcf})  yields the final expression for the signal
\begin{align}\label{eq:S1tsj}
&S^{(1)}(t,\omega)=-\mathcal{I}\frac{\mathcal{D}^2(\omega)}{\sigma_\omega}|\mu_{eg}|^2\rho_{ee}(0)\notag\\
&\times\left[\frac{1}{2\sigma_T\Delta_{eg}^-}+\frac{2i\Omega_1e^{-kt}}{(k-2i\Omega_1)(k+2\sigma_T)}\left(\frac{1}{\Delta_{eg}^-}-\frac{1}{\Delta_{eg}^+-ik}\right)\right],
\end{align}
where $\Delta_{eg}^{\pm}=\omega-\omega_{eg}^{\pm}-i(\sigma_T+\sigma_\omega)$. One can see clearly that in the short time limit the term with $d$ state: $\Delta_{eg}^-$ cancels out and the signal is dominated by the $u$ state via $\Delta_{eg}^{+}$. In the long time limit the second term in the square bracket is small and the signal is governed by a $d$ state via $\Delta_{eg}^-$.

The coincidence signal (\ref{eq:S2g}) is described by the four diagrams shown in Fig. \ref{fig:diags2}. After the change of variables it can be shown that diagrams $iii$ and $iv$ are complex conjugate of $ii$ and $i$, respectively. Diagram $i$ is analogous to nonrephasing, whereas diagram $ii$ corresponds to rephasing contributions to 2D optical signals. Reading the signal off the diagrams yields
\begin{align}
S{(2)}(t_1,\omega_1;t_2,\omega_2)&=\sum_{j=i}^{iv}S_j^{(2)}(t_1,\omega_1;t_2,\omega_2)\notag\\
&=2\mathcal{R}\sum_{j=i}^{ii}S_j^{(2)}(t_1,\omega_1;t_2,\omega_2),
\end{align}
where
\begin{align}\label{eq:S22i}
&S_i^{(2)}(t_1,\omega_1;t_2,\omega_2)=\mathcal{D}^2(\omega_1)\mathcal{D}^2(\omega_2)\hbar^4|\mu_{eg}|^2|\mu_{ef}|^2\notag\\
&\times\int dt_1' \int d\tau_1 D_>^{(1)}(t_1\omega_1;t_1',\tau_1)\notag\\
&\times\int dt_2' \int d\tau_2 D_>^{(2)}(t_2,\omega_2;t_2',\tau_2)\notag\\
&\times\langle \langle I|\mathcal{G}_{ge,ge}(\tau_1)\mathcal{G}_{ee,ee}(t_1'-t_2'-\tau_2)\mathcal{G}_{ef,ef}(\tau_2)\mathcal{G}_{ff,ff}(t_2')|\rho_{ff}\rangle\rangle_S,
\end{align}
\begin{align}\label{eq:S22ii}
&S_{ii}^{(2)}(t_1,\omega_1;t_2,\omega_2)=\mathcal{D}^2(\omega_1)\mathcal{D}^2(\omega_2)\hbar^4|\mu_{eg}|^2|\mu_{ef}|^2\notag\\
&\times\int dt_1' \int d\tau_1 D_>^{(1)}(t_1\omega_1;t_1',\tau_1)\notag\\
&\times\int dt_2' \int d\tau_2 D_<^{(2)}(t_2,\omega_2;t_2',\tau_2)\notag\\
&\times\langle \langle I|\mathcal{G}_{ge,ge}(\tau_1)\mathcal{G}_{ee,ee}(t_1'-t_2')\mathcal{G}_{fe,fe}(-\tau_2)\mathcal{G}_{ff,ff}(t_2'+\tau_2)|\rho_{ff}\rangle\rangle_S.
\end{align}
Here $D_>$ ($D_<$) indicate the sign of $\tau$ and corresponds to the first (second) term in the square bracket in Eq. (\ref{eq:WDlor}). The initial state in this case is the spin up state $|\rho_{ff}\rangle\rangle_S=\rho_{ff}(0)|ff\rangle\rangle\begin{pmatrix}
  1 \\
  0 \\
 \end{pmatrix}$. The final form for the signal is given in (\ref{eq:S23i}) - (\ref{eq:S23ii}).
Careful analysis shows that for short $t_2$ the PCC signal is dominated by $\Delta_{fe}^{+}$ and for long $t_2$ it is dominated by $\Delta_{fe}^{-}$ in spectral gate of the second detector $\omega_2$. Similarly for short $t_1-t_2$ the signal is governed by $\Delta_{eg}^{+}$ components whereas for long $t_1-t_2$ it is dominated by $\Delta_{eg}^{-}$ for the spectral gate of the first detector $\omega_1$.

\section{Simulations}\label{sec:Sim}

Below we simulate the first order -   $S^{(1)}(t,\omega)$ Eq. (\ref{eq:S1tsj}) and second order - $S^{(2)}(t_1,\omega_1;t_2,\omega_2)$ photon counting signals given by  Eqs. (\ref{eq:S23i}) - (\ref{eq:S23ii}).  Simulation parameters are  $\Omega_0=12500$ cm$^{-1}$, $\Omega_1=125$ cm$^{-1}$, $\Delta_0=250$ cm$^{-1}$, $\Delta_1=5$ cm$^{-1}$, dephasings $\gamma_e=8.56$ cm$^{-1}$, $\gamma_f=17.22$ cm$^{-1}$, jump rate $k=7.68$ cm$^{-1}$, dipole moments $\mu_{fe}=\sqrt{2}\mu_{eg}$.

\begin{figure}[t]
\begin{center}
\includegraphics[trim=0cm 0cm 0cm 0cm,angle=0, width=0.45\textwidth]{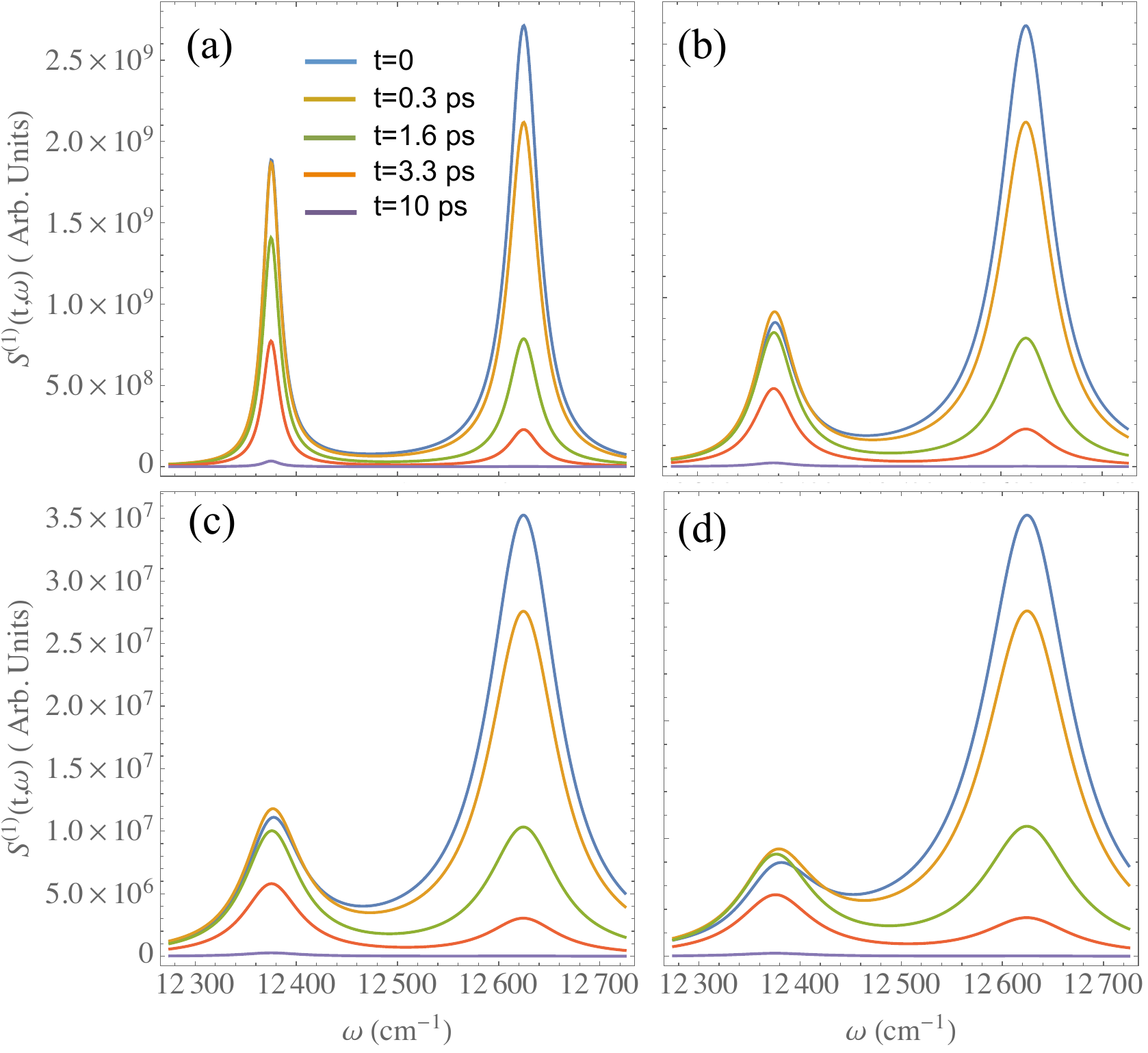}
\end{center}
\caption{(Color online) The first order photon counting signal $S^{(1)}(t,\omega)$ (\ref{eq:S1tsj}) for the TSJ model as a series of snapshots at different time $t$. Different panels correspond to different gating bandwidths: (a) - $\sigma_T=0.7$ cm$^{-1}$, $\sigma_\omega=0.8$ cm$^{-1}$, (b) - $\sigma_T=7$ cm$^{-1}$, $\sigma_\omega=8$ cm$^{-1}$, (c) - $\sigma_T=7$ cm$^{-1}$, $\sigma_\omega=18$ cm$^{-1}$, (d) - $\sigma_T=17$ cm$^{-1}$, $\sigma_\omega=18$ cm$^{-1}$. All other parameters is listed in Section \ref{sec:Sim}.}
\label{fig:S1gate}
\end{figure}

\begin{figure}[t]
\begin{center}
\includegraphics[trim=0cm 0cm 0cm 0cm,angle=0, width=0.45\textwidth]{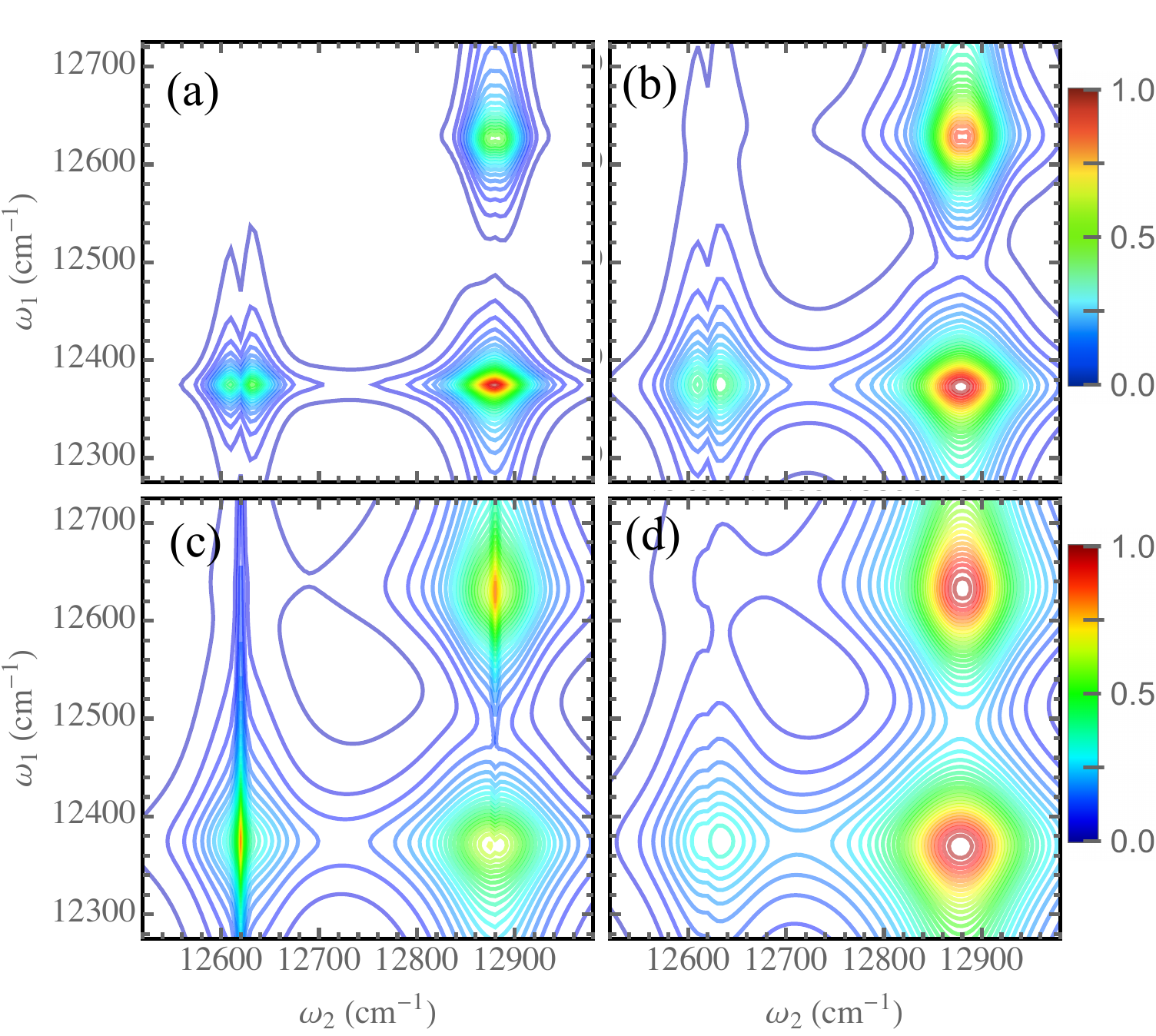}
\end{center}
\caption{(Color online) The 2D second order photon counting signal $S^{(2)}(t_1,\omega_1;t_2,\omega_2)$ for the TSJ model (\ref{eq:S23i}) - (\ref{eq:S23ii}) as 2D frequency correlation plot  vs $\omega_1$ and $\omega_2$ for fixed times $t_2=3.3$ ps, $t_1-t_2=3.3$ fs. Different panels correspond to different gating bandwidths: (a) - $\sigma_{T1}=0.7$ cm$^{-1}$, $\sigma_{\omega1}=0.8$ cm$^{-1}$, $\sigma_{T2}=0.75$ cm$^{-1}$, $\sigma_{\omega2}=0.85$ cm$^{-1}$ (b) - $\sigma_{T1}=7$ cm$^{-1}$, $\sigma_{\omega1}=8$ cm$^{-1}$, $\sigma_{T2}=7.5$ cm$^{-1}$, $\sigma_{\omega2}=8.5$ cm$^{-1}$, (c) - $\sigma_{T1}=7$ cm$^{-1}$, $\sigma_{\omega1}=18$ cm$^{-1}$, $\sigma_{T2}=7.5$ cm$^{-1}$, $\sigma_{\omega2}=18.5$ cm$^{-1}$,  (d) - $\sigma_{T1}=17$ cm$^{-1}$, $\sigma_{\omega1}=18$ cm$^{-1}$, $\sigma_{T2}=17.5$ cm$^{-1}$, $\sigma_{\omega2}=18.5$ cm$^{-1}$. All other parameters is listed in Section \ref{sec:Sim}.}
\label{fig:S2gate}
\end{figure}

\begin{figure*}[t]
\begin{center}
\includegraphics[trim=0cm 0cm 0cm 0cm,angle=0, width=0.95\textwidth]{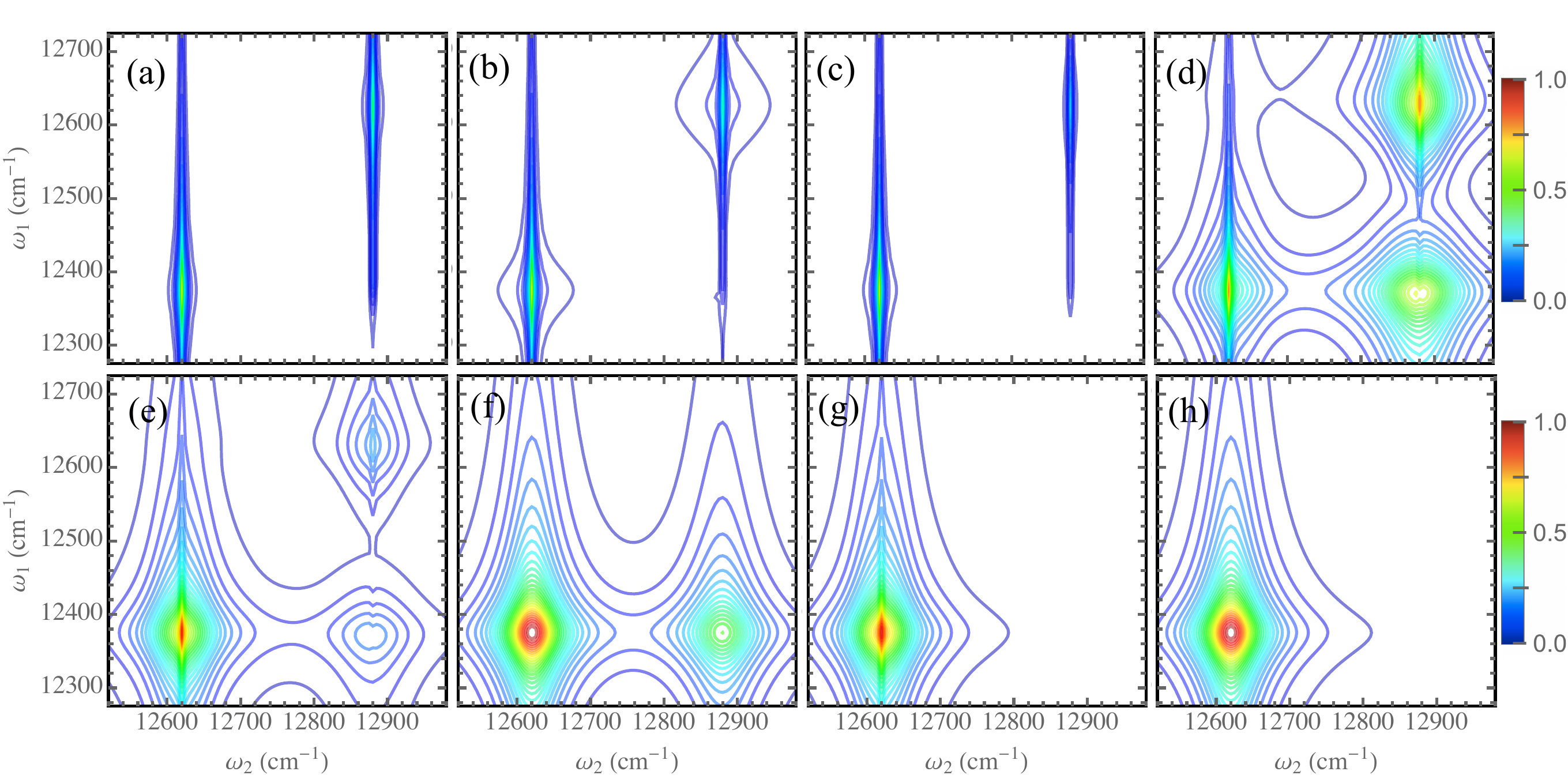}
\end{center}
\caption{(Color online) The 2D second order photon counting signal $S^{(2)}(t_1,\omega_1;t_2,\omega_2)$ (\ref{eq:S23i}) - (\ref{eq:S23ii}) as 2D frequency correlation plot  vs $\omega_1$ and $\omega_2$ at fixed gating parameters from Fig. \ref{fig:S2gate}c. Different panels correspond to various values of the times $t_1$ and $t_2$: (a) - $t_2=t_1-t_2=3.3$ fs, (b) - $t_2=t_1-t_2=1$ ps, (c) - $t_2=3.3$ ps, $t_1-t_2=3.3$ fs, (d) - $t_2=3.3$ fs, $t_1-t_2=3.3$ ps, (e) - $t_2=t_1-t_2=3.3$ ps, (f) - $t_2=3.3$ ps, $t_1-t_2=33$ ps, (g) $t_2=33$ ps, $t_1-t_2=3.3$ ps, (h)  - $t_2=t_1-t_2=33$ ps. The rest of the parameters is listed in Section \ref{sec:Sim}.}
\label{fig:S2tf}
\end{figure*}

\begin{figure*}[t]
\begin{center}
\includegraphics[trim=0cm 0cm 0cm 0cm,angle=0, width=0.95\textwidth]{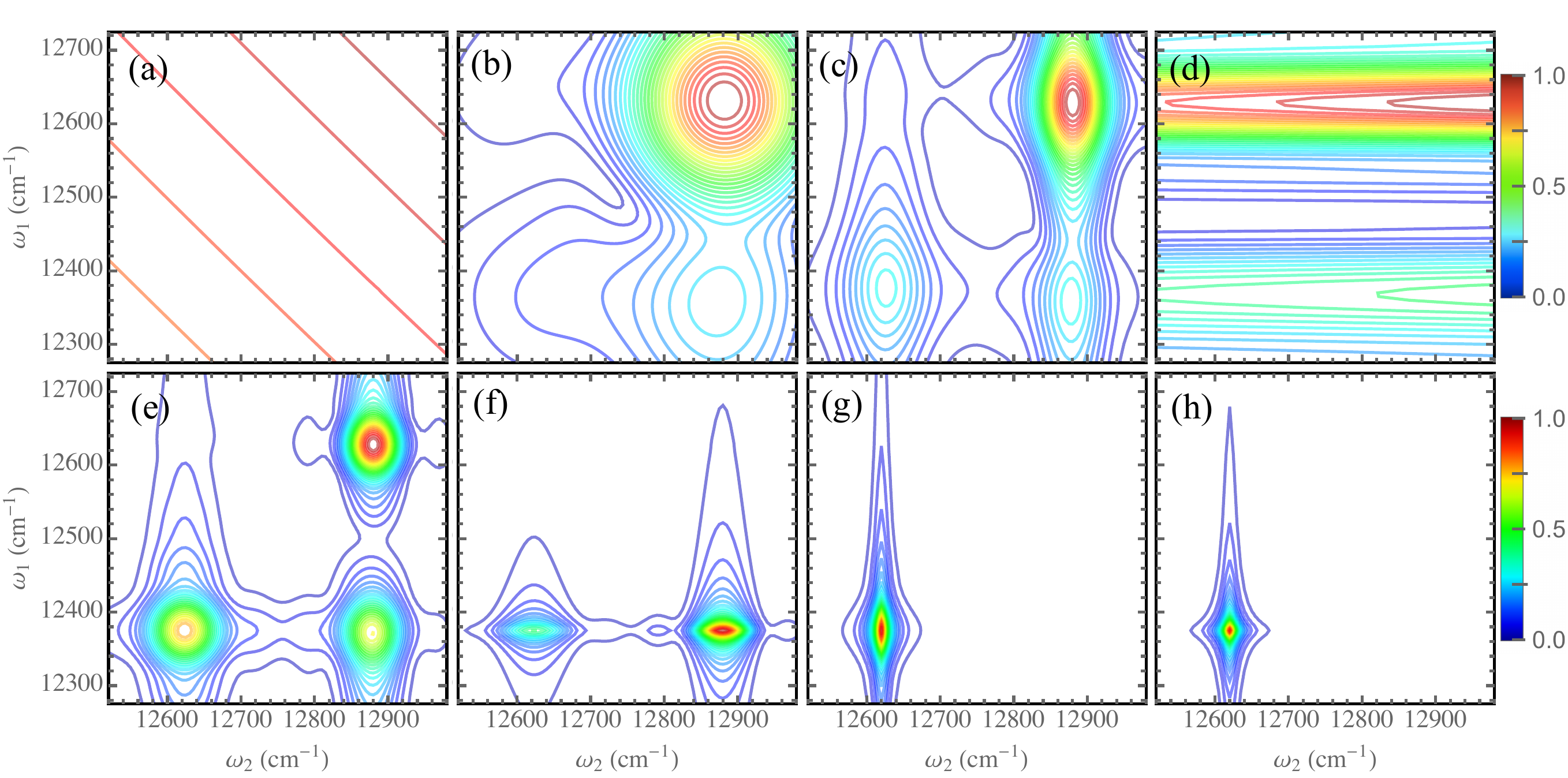}
\end{center}
\caption{(Color online) The 2D second order photon counting signal $S^{(2)}(t_1,\omega_1;t_2,\omega_2)$ for the TSJ model calculated using physical spectrum for the same parameters as in Fig. \ref{fig:S2tf}.}
\label{fig:S2ps}
\end{figure*}

\begin{figure}[t]
\begin{center}
\includegraphics[trim=0cm 0cm 0cm 0cm,angle=0, width=0.45\textwidth]{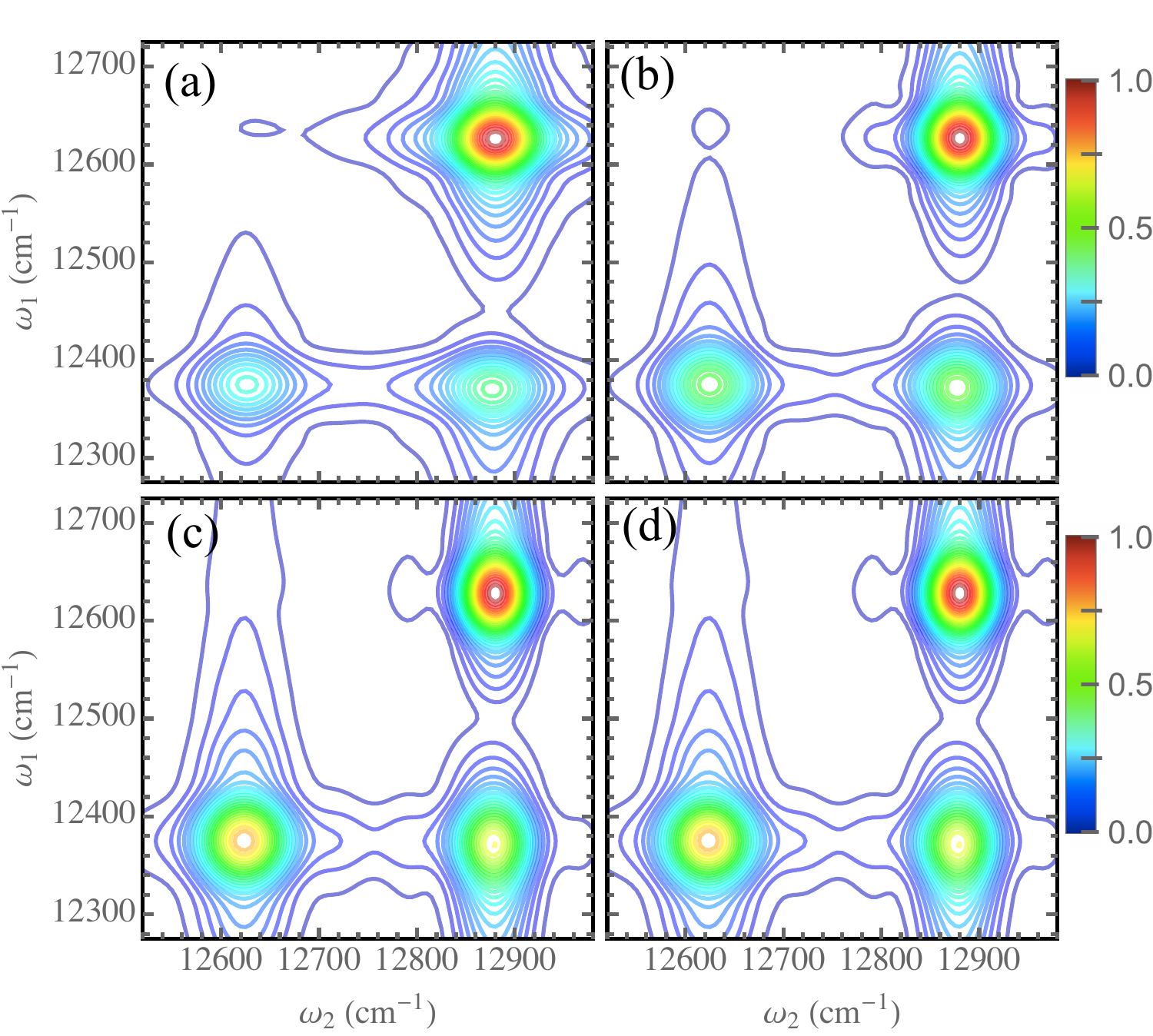}
\end{center}
\caption{(Color online) The 2D second order photon counting signal $S^{(2)}(t_1,\omega_1;t_2,\omega_2)$ for the TSJ model calculated using physical spectrum for fixed times $t_2=3.3$ ps, $t_1-t_2=3.3$ ps.  Panels a-d  and the rest of the parameters are the same as those in Fig. \ref{fig:S2gate}.}
\label{fig:S2psgate}
\end{figure}

Fig. \ref{fig:S1gate} illustrates the effect of gating on  $S^{(1)}(t,\omega)$ . For narrow time and frequency gates Fig. \ref{fig:S1gate}a both up and down frequencies $\omega_{eg}^{\pm}$ show up as narrow resonances dominated at short time by $\omega_{eg}^{+}$ and at longer time - by $\omega_{eg}^{-}$. The overall magnitude of the signal decreases with time, and due to the narrow time gate this decrease is noticeable between the spectra at different $t$. When both time and frequency bandwidths of the gates are increased Fig. \ref{fig:S1gate}b the overall spectral resolution is suppressed and the resonances broaden. The time resolution decreases as well, so the distinction between signals at maximum resonances become less pronounced.  When further increasing the spectral gate bandwidth while keeping the time gate intact, Fig. \ref{fig:S1gate}c shows that only the spectral broadening increases, whereas the distinction in time remain the same as in panel $b$. Finally when both time and frequency gates become more wide than the jump rate $k$, both spectral and temporal resolutions decrease and the signal is further broadens and distinction at different times becomes even less apparent.

We next turn to gated second order counting signal  $S^{(2)}(t_1,\omega_1;t_2,\omega_2)$ . Fig. \ref{fig:S2gate}a shows three resonant peaks. As in the case of the photon echo - two diagonal peaks correspond to populations and are governed by two uncoupled cascading and $f\to e$ and $e\to g$ transitions. These two peaks occur at $\omega_2=\omega_{fe}^{\pm}$ but are dominated by $\omega_{fe}^{+}$. An additional off-diagonal peak at $\omega_1=\omega_{eg}^{+}$, $\omega_2=\omega_{fe}^{+}$ is governed by the coherence induced by jump which couples two cascading transitions.  These three peaks can be explained in the following way. At short time $t_2$ the peak at $\omega_2=\omega_{fe}^{+}$ is much stronger than at $\omega_2=\omega_{fe}^{-}$. Similarly at long $t_1-t_2$  most of the population is at $\omega_1=\omega_{eg}^{-}$. The seemingly doubling of the resonance at $\omega_2=\omega_{fe}^{-}$ is due to the oscillating phase of the signal that depends on the $\omega_2$ and becomes narrower than the chosen gating bandwidth. An interesting effect occurs by increasing the time and frequency resolution. Fig. \ref{fig:S2gate}b shows that besides obvious decrease in spectral resolution, which results in overall broadening of the peaks, the intensity distribution between three peaks change as well. In panel $a$ we saw that most population is concentrated at one peak at $\omega_2=\omega_{fe}^{+}$ and $\omega_1=\omega_{eg}^{-}$. Here we see that the peak at $\omega_2=\omega_{fe}^{+}$ and $\omega_1=\omega_{eg}^{+}$ becomes stronger as well but still weaker than the former.  This occurs since the broader time gate bandwidth $\sigma_{T1}$ allows to capture the population in  $\omega_{eg}^{+}$ transition that was present there at earlier times and is contributing now to the signal. Fig. \ref{fig:S2gate}c shows that if we increase only the spectral gate, while keeping the time gate the same as in panel $b$, the overall spectral resolution for detector $1$ is decreased, whereas it becomes better for detector $2$. This interesting effect occurs, since the actual spectral resolution for $\omega_2$ is a combination of the bandwidths of both detectors 1 and 2, and the combined bandwidth enters in different terms of the signal with different signs. Some terms depend on the sum of the both detector's time and frequency bandwidths, some - depend on their difference. For certain parameter regime, the increase in the gating bandwidth can actually help to resolve the peaks better. Finally when both time and frequency gating bandwidths become broad in both detectors, Fig. \ref{fig:S2gate}d shows that the overall resolution drops, while the relative intensity between two intense peaks at $\omega_1=\omega_{eg}^{\pm}$ become equal due to the larger value of the $\sigma_{T1}$.

We next present frequency-frequency correlation spectra of $S^{(2)}(t_1,\omega_1;t_2,\omega_2)$ at fixed times $t_1$ and $t_2$. We first fix the gating bandwidth as in the case of Fig. \ref{fig:S2gate}c. Fig. \ref{fig:S2tf}a shows that if both $t_1$ and delay $t_2-t_1$ are short (much shorter than the jump rate $k$), the signal shows two diagonal peaks at $\omega_2=\omega_{fe}^{+}$, $\omega_1=\omega_{eg}^{+}$ and $\omega_2=\omega_{fe}^{-}$, $\omega_1=\omega_{eg}^{+}$ which are due to uncoupled cascading transitions that don't give rise to the coherence since the temporal gates are much narrower than the jump rate and dephasing and the short time dynamics is governed by these uncoupled transition. Further increase of the time delays $t_2$ and $t_1-t_2$ shown in panel b shows a slight increase in  the peak intensities due to larger time, whereas panel $c$ shows a decrease in the intensity of the peak at $\omega_1=\omega_{eg}^{+}$ and $\omega_2=\omega_{fe}^{-}$ due to smaller delay time $t_2$. Panel $d$ shows a new third off-diagonal peak since the delay $t_1-t_2$ becomes large enough to accommodate the $\omega_{eg}^{-}$ transition. Broader resonances are seen in panel $d$. When both $t_2$ and $t_1-t_2$ become relatively large (around $3.3$ ps) the intensity of the three peaks becomes redistributed towards the lowest peak $\omega_{eg}^{-}$ and $\omega_{fe}^{-}$ as shown in Fig. \ref{fig:S2tf}e. Further increase of the $t_1-t_2$ eliminates the higher energy diagonal peak at $\omega_1=\omega_{eg}^{+}$ as shown in Fig. \ref{fig:S2tf}f. Finally when $t_2$ becomes large as well Fig. \ref{fig:S2tf}g, and h show that the peak at $\omega_{fe}^{+}$ is eliminated as well leaving only the ``down'' states contribution via a single diagonal peak at $\omega_1=\omega_{eg}^{-}$ and $\omega_2=\omega_{fe}^{-}$ as can be expected from the TSJ model.

We now compare Fig. \ref{fig:S2tf} with the ``physical spectrum''. This is done by replacing the time-and-frequency resolved detector spectrograms in Eqs. (\ref{eq:S22i}) - (\ref{eq:S22ii}) by the physical spectrum  $D_>(t,\omega;t',\tau)\to D_{ps}(t,\omega;t',\tau)$ where
\begin{align}
D_{ps}(t,\omega;t',\tau)&=\theta(t-t')\theta(t-t'-\tau)\notag\\
&\times e^{(i\omega-\Gamma/2)(t-t')-(i\omega+\Gamma/2)(t-t'-\tau)}.
\end{align}
We took $\Gamma = \sigma_\omega$ as this is a pure frequency gate, so all the rest of the parameters are exactly the same as in Fig. \ref{fig:S2tf}. Comparison of Fig. \ref{fig:S2ps} with Fig. \ref{fig:S2tf} shows that for physical spectrum is unable to capture the system dynamics at short times (top row - panels a-d), especially at short $t_2$ time (panels a and d) which does not give any spectral resolution for detector 2. In the same time, the physical spectrum gives very well resolved long time dynamics which is depicted by panels (e-h). This is a consequence of the fact that physical spectrum is obtained using a time-stationary filter, whereas the time-and-frequency resolved gating uses time-nonstationary filters. Clearly the characterization of pulsed experiments requires time-nonstationary filters as shown in \cite{iac98}. Moreover, the physical spectrum at long time delays $t_2$ and $t_1-t_2$ has higher resolution than the time-and-frequency resolved measurement shown in Fig. \ref{fig:S2tf}. This is due to the fact that physical spectrum is governed by a single parameter of the frequency gate bandwidth which enters as the peak spectra width. Time-and-frequency resolved measurements have a linear combination of the time and frequency gate which governs the linewidth and can be quite broad if all the bandwidths add up. Another important distinction of the physical spectrum is that for given time delays $t_2$ and $t_1-t_2$ the physical spectrum shows a different intensity profile for the same resonances compared to the time-and-frequency resolved measurement. This is a consequence of the better control over the temporal and spectral bandwidths of the signal. Larger bandwidths allow to capture photons at earlier times which may come from an up- states.

Another important feature of the physical spectrum is that it does not correctly include the quantum fluctuations in the higher order field correlation measurement. A typical example of such a spectrum is the squeezing spectrum of the atomic resonance fluorescence. Its first treatment led to correct results \cite{col84}. However, in that treatment the authors used  inconsistent spectral filtering and ignored time-delayed commutators. As discussed in Ref. \cite{kno86}, these commutator terms are not negligible and they  are not properly incorporated in the filtering theory. Using the present superoperator approach formulated in the joint field plus matter space analysis we have earlier demonstrated that commutator terms are  necessary for computing the relevant components of the nonlinear susceptibility in the parametric down conversion generation \cite{dor121}. Furthermore, this approach allows to express all the correct features of the multipoint field correlation functions without adding phenomenological quantum noise terms that are typically needed to account for higher order correlation of the multimode quantum states of light.

Finally, the control of the temporal resolution of the physical spectrum is relatively low. We already discussed in detail Fig. \ref{fig:S2gate} where we showed how by changing gating bandwidths one can significantly change the intensity of various peaks. Fig. \ref{fig:S2psgate} shows the corresponding physical spectrum result. One can see that the resonance at $\omega_1=\omega_{eg}^{+}$, $\omega_2=\omega_{fe}^{+}$ remain the strongest for all the values of the gating bandwidths. Two peaks at $\omega_1=\omega_{eg}^{-}$, $\omega_2=\omega_{fe}^{\pm}$ have roughly equal intensity and are weak. Increasing the bandwidths of the gates slightly increase their intensity from panel a to d.

\section{Discussion}\label{sec:conclusion}

Eberly and Wodkiewicz \cite{Ebe77} had argued that specific detector gating with finite bandwidth must be added to describe the real detector. A two-level detector characterized by a single parameter $\Gamma$ that accounts for both time-and-frequency detection  is known as the  physical spectrum \cite{del12,gon15}. This result  can be recovered from our model  by simply removing the time gate $F_t=1$ and only a Lorentzian frequency gate such that
\begin{align}
F_f(\omega,\omega')=\frac{i}{\omega'+\omega+i\Gamma/2}.
\end{align}
Using the physical spectrum time-and-frequency resolved photon coincidence signal is given by
\begin{align}\label{eq:g2dV}
&g_{\Gamma_1\Gamma_2}^{(2)}(\omega_1,\omega_2;\tau)=\notag\\
&\text{lim}_{t\to\infty}\frac{\langle \hat{A}_{\omega_1,\Gamma_1}^{\dagger}(t)\hat{A}_{\omega_2,\Gamma_2}^{\dagger}(t+\tau)A_{\omega_2,\Gamma_2}(t+\tau)\hat{A}_{\omega_1,\Gamma_1}(t)\rangle}{\langle \hat{A}_{\omega_1,\Gamma_1}^{\dagger}(t)\hat{A}_{\omega_1,\Gamma_1}(t)\rangle\langle \hat{A}_{\omega_2,\Gamma_2}^{\dagger}(t+\tau)\hat{A}_{\omega_2,\Gamma_2}(t+\tau)\rangle},
\end{align}
where
\begin{align}\label{eq:gdV}
\hat{A}_{\omega,\Gamma}(t)=\int_{-\infty}^tdt_1e^{(i\omega-\Gamma/2)(t-t_1)}\hat{E}(t_1),
\end{align}
is a gated field. This model which provides a simple benchmark for finite-band detector analysis, has several limitations. First, the parameters of the time and frequency gates are not independent, this cannot fully represent experimental setups where frequency filter and avalanche photodiode are two independent devices. Second, this method does not fully address the generation and photon bandwidth coming from the emitter, as the analysis is performed in the field space alone. Finally, the multi photon correlation function presented in \cite{gon15} is stationary. For instance, the four-point  bare correlation function
\begin{align}\label{eq:bare0}
A_B(\omega_1,\omega_2,t_1,t_2)=\langle E_{\omega_1}^{\dagger}(t_1)E_{\omega_2}^{\dagger}(t_2)E_{\omega_2}(t_2)E_{\omega_1}(t_1)\rangle
\end{align}
depends on four times and four frequencies. After gating  \cite{del12,gon15} the correlation function (\ref{eq:bare0}) becomes dependent upon $C_B(\omega_1,\omega_2,t_2-t_1)$, which depends only on the time difference $t_2-t_1$, which is approximation for a stationary fields. This model also works only  if $t\gg\Gamma^{-1}$, which means that $\Gamma$ cannot approach zero (e.g.perfect reflection in Fabri Perot cavity). It also works when $\Gamma\tau_0\ll1$ where $\tau_0$ is the scale of change in the field envelope. For comparison, the photon coincidence counting  (PCC) (\ref{eq:gN}) for $N=2$ reads
\begin{align}\label{eq:g2}
&g^{(2)}(t_1,\omega_1,\Gamma_1;t_2;\omega_2,\Gamma_2)\frac{\langle \hat{n}_{t_1,\omega_1}\hat{n}_{t_2,\omega_2}\rangle}{\langle\hat{n}_{t_1,\omega_1}\rangle\langle\hat{n}_{t_2,\omega_2}\rangle},
\end{align}
which depends on two times $t_1$, $t_2$ and two frequency $\omega_1$, $\omega_2$ arguments.

Eq. (\ref{eq:g2}) has several important advantages compared to theory of \cite{del12,gon15}. First,  independent control of time and frequency gates (with guaranteed Fourier uncertainty for the time and frequency resolution) along with the fact that bare photon number operator depends on two time variables $\hat{n}(t,\tau)$ allows to capture any dynamical process down to very short scale dynamics.  Second, the gating (\ref{eq:nDn}) provide a versatile tool that can capture nonequlibrium and non stationary states of matter which can be controlled by gating bandwidths. In this case a series of frequency correlation plots for $\omega_1$, $\omega_2$ (keeping the central frequencies of the spectral gates as variables) for different time delays $t_1-t_2$ yields a 2D spectroscopy tool capable of measuring ultrafast dynamics. Third, superoperator algebra allows to connect  the gated field correlation function (\ref{eq:S20}) with the bare correlation function (\ref{eq:bare0}), using arbitrary time-and-frequency gates (not necessarily Lorentzian) as well as material response that precedes the emission and detection of photons. The superoperator expressions require time ordering and therefore can be generalized on other correlation functions of the field operators that are not normally ordered. Superoperator algebra is an effective tool for bookkeeping field-matter interactions prior to the spontaneous emission of photons. Finally, as we show in the next section PCC can be recast in terms of matter correlation function by expanding the total density matrix operator in perturbation series and tracing the vacuum modes. Photon counting measurement can thus be related to matter response as is done in nonlinear spectroscopy.

In summary, time-and-frequency gated photon counting may provide a  novel tool for multidimensional optical spectroscopy. Besides the low intensity requirement that allows to have low background noise and avoid laser intensity fluctuations, photon counting provides similar and complementary information to coherent multidimensional techniques. In particular the time-and-frequency coincidence signals provide information similar to that of a 2D photon echo experiments (rephasing and nonrephasing contributions).  The independent variation of the time and frequency gating parameters allow to have extra control knobs for manipulation of the signals. This technique improves the conventional physical spectrum in several ways. First, it allows to observe the short time system dynamics. Second, the time and frequency resolution are controlled independently, whereas in the physical spectrum both are determined by a single parameter. Finally the  microscopic origin of the proposed method allows to connect photon counting observable with the multipoint matter correlation functions that govern the system properties. We note that the time-and-frequency resolved detection requires ultrafast time gating which is not trivial. However recent advances in ultrafast upconversion  techniques for the single photon detectors can resolve this problem by allowing to time the photons with up to $\sim100$ fs resolution \cite{Kuz08}.

\acknowledgements
We gratefully acknowledge the support of the National Science Foundation through Grant No. CHE- 1361516, the Chemical Sciences, Geosciences and Biosciences Division, Office of Basic Energy Sciences, Office of Science, U.S. Department of Energy Award \#DE-FG02-04ER15571. KD was supported by the DOE which also provided the computational resources. We wish to thank Dr. Frank Schlawin for useful discussions.


\appendix

\section{Gated photon counting measurements}\label{app:field}

An $N$-th order photon correlation measurement is given by the following normally-ordered product of the electric field operators \cite{Glau07}
\begin{align}\label{eq:gN0}
&g^{(N)}=
\frac{:\langle \hat{n}_{1}...\hat{n}_{N}\rangle:}{\langle\hat{n}_{1}\rangle...\langle \hat{n}_{N}\rangle},
\end{align}
where $\hat{n}_j=\hat{E}_j^{\dagger}\hat{E}_j$ is the photon number and $:\langle...\rangle :$ denotes the normally ordered combination of operators. The correlation function (\ref{eq:gN0}) is measured in a $N$-th coincidence measurement using $N$ photon detectors. In order to extract physical parameters of light such as photon statistics, correlations etc. one has to understand how the photon detector works. To a good approximation  we can represent an ideal detector by two-level atom that is initially in the ground state $b$ and is promoted to the ionization continuum represented by the manifold $a$ by the absorption of a photon (see Fig. \ref{fig:diags1}a). Photon detection brings the field from its initial state $\psi_i$ to a final state $\psi_f$. The probability amplitude for photon absorption at time $t$ can be calculated in first-order perturbation theory \cite{Glau07}
\begin{equation}
\langle\psi_f|\hat{\mathbf{E}}(t)|\psi_i\rangle\cdot\langle a|\mathbf{d}|b\rangle,
\end{equation}
where $\mathbf{d}$ is the dipole moment of the atom and $\hat{\mathbf{E}}(t)=\hat{E}^{\dagger}(t)+\hat{E}(t)$ is the electric field operator (we omit the spatial dependence). Clearly, only the annihilation part of the electric field contributes to the photon absorption process. The transition probability to find the field in state $\psi_f$ at time $t$ is given by the modulus square of the transition amplitude
\begin{align}\label{eq:Glau}
&\sum_{\psi_f}|\langle\psi_f|\hat{E}(t)|\psi_i\rangle|^2=\langle\psi_i|
\hat{E}^{\dagger}(t)\sum_{\psi_f}|\psi_f\rangle\langle\psi_f|\hat{E}(t)|\psi_i\rangle\notag\\
&=\langle\psi_i|\hat{n}(t)|\psi_i\rangle.
\end{align}
Since the initial state of the field $\psi_i$ is rarely known with certainty, we must trace over all possible initial states as determined by the field density operator $\rho$. Thus, the output of an ideal detector is more generally given by $\text{tr}\left[\rho E^{\dagger}(t)E(t)\right]$. Eq. (\ref{eq:Glau}) represents an ideal detector, which has infinite spectral bandwidth.  Real detectors have finite temporal and spectral resolutions controlled by a shutter, streak camera, or avalanche photodiode in case of temporal resolution and spectrometer in the case of spectral resolution. One can think of a realistic detector as a sequence of time and frequency gates that acquire  temporal and spectral information about the detected photons. In many applications of quantum technologies, such as lithography, or quantum optics the temporal resolution is not crucial and the temporal gate is removed. Here we include both spectral and temporal gates. The gating works in the following way. First a time gate governed by a $F_t(t',t)$ centered at time $t$ is applied to a bare field (undated) $\hat{E}(t')$:
\begin{align}\label{eq:Et}
\hat{E}_t(t')=F_t(t',t)\hat{E}(t').
\end{align}
In order for $F_t(t',t)$ to represent unitary transformation one has to add Langevin noise operator \cite{vog06}. This is done to account for vacuum fluctuations caused by other field modes. However, we have recently shown \cite{dor121} how the multipoint correlation function of the field is calculated in the joint field plus matter space given by the sum over Liouville space pathways. This microscopic method can account for all the effects of multiple vacuum modes by theses paths without adding a Langevin noise source.
We then apply a spectral gate governed by a function $F_f(\omega',\omega)$ centered at frequency $\omega$ is applied to spectral component of time-gated field (\ref{eq:Et}) $\hat{E}_t(\omega',\omega)$
\begin{align}
\hat{E}_{t,\omega}(\omega')=F_f(\omega',\omega)\hat{E}_t(\omega').
\end{align}
Finally the time-and-frequency gated field operator $\hat{E}_{t,\omega}(t'')$  is defined in terms of the bare field operator $\hat{E}(t)$ as follows
\begin{align}\label{eq:eft}
&\hat{E}_{t,\omega}(t'')=\int_{-\infty}^{\infty}dt'F_f(t''-t',\omega)F_t(t',t)\hat{E}(t').
\end{align}
Alternatively if the frequency gate is applied before the time gate, the gated field operator is given by
\begin{align}\label{eq:etf}
&\hat{E}_{\omega, t}(t'')=\int_{-\infty}^{\infty}dt'F_t(t'',t)F_f(t''-t',\omega)\hat{E}(t').
\end{align}
In the following we will employ the protocol of (\ref{eq:eft}). Eq. (\ref{eq:etf}) can be done similarly.

We now turn to the gated photon counting signals. This will be done by working in Liouville space \cite{Mukamel_book, Nagata11}, $i.e.$ the space of bounded operators in the joint matter-field Hilbert space. It offers a convenient bookkeeping device for matter-light interactions. Signals are described as time-ordered products of superoperators. We first introduce the basic notation. With each Hilbert space operator $A$ we associate two superoperators \cite{Harbola08a}
\begin{align}
A_L X &\equiv A X, \label{eq.A_L}
\end{align}
which represents the action from the left, and
\begin{align}
A_R X &\equiv X A, \label{eq.A_R}
\end{align}
which implies action from the right. We further introduce two linear combinations of the left and right operators, the commutator superoperator
\begin{align}
A_- &\equiv A_L - A_R, \label{eq.commutator}
\end{align}
and the anti-commutator
\begin{align}
A_+  &\equiv \frac{1}{2} \left( A_L + A_R \right). \label{eq.anti-commutator}
\end{align}
This notation allows to derive compact expressions for spectroscopic signals. At the end of the calculation, after the time-ordering is taken care of, we can switch back to ordinary Hilbert space operators. 


\section{Derivation of the matter correlation expressions for the gated single photon counting signals}\label{app:S1}

The lowest order signal related to photon counting is time-and-frequency resolved photon number
\begin{align}\label{eq:S10}
S^{(1)}(t,\omega)\equiv n_{t,\omega}=\int dt'' \langle\hat{E}_{t,\omega R}^{\dagger}(t'')\hat{E}_{t,\omega L}(t'')\rangle_T,
\end{align}
where the integration over $t''$  collects signal for extended period of time limited by the temporal bandwidth of the time gate.  $\langle ...\rangle_T=\text{Tr}[...\rho_T(t)]$ and 
\begin{align}\label{eq:rhoT}
\rho_T(t)=\mathcal{T} e^{-\frac{i}{\hbar}\int d\tau \hat{H}_-'(\tau)},
\end{align}
represents the total field plus matter density matrix. It contains information about system evolution prior to the detection (e.g. photon generation process, etc.), $\mathcal{T}$ is a time-ordering superoperator that rearranges products of superoperators such that their time arguments increase from right to left. The field-matter interaction Hamiltonian in the interaction picture and the rotating-wave approximation reads
\begin{align}
\hat{H}_-'(t)=\hat{V}_L(t)\hat{E}_L^{\dagger}(t)-\hat{V}_R(t)\hat{E}_R^{\dagger}(t)+H.c.,
\end{align}
where $\hat{V}(\hat{V}^{\dagger})$ are lowering (raising) dipole operator $\hat{V}=\mu\hat{\sigma}_-$, $\hat{V}^{\dagger}=\mu^{*}\hat{\sigma}_+$ given by the ladder matrices $\hat{\sigma}_{\pm}=\hat{\sigma}_x\pm i\hat{\sigma}_y$ where $\hat{\sigma}_x$ and $\hat{\sigma}_y$ are Pauli matrices and $\mu$ is the transition dipole moment.  Note, that unlike the standard notation in (\ref{eq:gN0}) there is no need to introduce normal ordering of operators; a time-ordering of the superoperators takes care of the book keeping.

Using the gating transformation (\ref{eq:eft}) one can recast Eq. (\ref{eq:S10}) as
\begin{align}\label{eq:nDn}
S^{(1)}(t,\omega)=\int dt' \int d\tau D(t,\omega;t',\tau)n(t',\tau).
\end{align}
Here
\begin{align}\label{eq:Ddef}
&D(t,\omega,t',\tau)=\notag\\
&\int\frac{d\omega''}{2\pi}e^{-i\omega''\tau}|F_f(\omega'',\omega)|^2F_t^{*}(t'+\tau,t)F_t(t',t).
\end{align}
$n(t',\tau)$ is a bare photon number defined in terms of superoperators as follows
\begin{align}\label{eq:ntt}
n(t',\tau)= \langle \hat{n}(t',\tau)\rangle_T,
\end{align}
where
\begin{align}
\hat{n}(t',\tau)=\sum_{s,s'}\hat{E}_{sR}^{\dagger}(t'+\tau)\hat{E}_{s'L}(t').
\end{align}
Here $\hat{E}_s(t)=\sqrt{2\pi\hbar\omega_s/\Omega}\hat{a}_se^{-i\omega_s t}$ and $\Omega$ is the mode quantization volume.

To clarify the implications of gating on the detected signal (\ref{eq:nDn}) we first consider the bare signal(\ref{eq:ntt}) using the loop diagram shown in Fig. \ref{fig:diags1}b.  The bare signal is given by the product of two transition amplitude superoperators \cite{Rah10} (one for bra and one for ket of the matter, and the field joint density matrix), each creating a coherence in the field between states with zero and one photon. By combining the transition amplitude superoperators from both branches of the loop diagram we obtain the photon occupation number that can be detected. Ideal frequency domain detection only requires  a single field mode \cite{Muk11}. However, maintaining any level of time resolution requires a superposition of several field modes. The underlying matter pathway information is not directly accessible in the standard detection theory that operates in the field space alone \cite{Glau07} and requires to work in the joint matter and field space. 
The leading contribution to $S^{(1)}$ signal (\ref{eq:nDn}) comes from second order expansion of field matter interactions with vacuum modes (see diagram in Fig. \ref{fig:diags1}b)
\begin{align}
n(t',\tau)&=\frac{1}{\hbar^2}\int_{-\infty}^{t'}dt_1\int_{-\infty}^{t'+\tau}dt_2\langle \mathcal{T}V_R^{\dagger}(t_2)\langle V_L(t_1)\rangle\notag\\
&\times\sum_{s,s'}\langle \mathcal{T}\hat{E}_{s'R}(t_2)\hat{E}_{s'R}^{\dagger}(t'+\tau)\hat{E}_{sL}(t')\hat{E}_{sL}^{\dagger}(t_1)\rangle_v,
\end{align}
where we utilized superoperator time ordering and $\langle ...\rangle=\text{Tr}[...\rho(t)]$ where $\rho(t)$ is the density operator that excludes vacuum modes and $\langle ...\rangle_v=\text{Tr}[...\rho_v(t)]$ where $\rho_v(t)=|0\rangle\langle 0|$ is the density matrix of the vacuum modes. One can now evaluate explicitly the vacuum field correlation function where $\hat{a}_s^{\dagger}(\hat{a}_s)$ is creation (annihilation) operators for mode s which satisfy boson commutation relation $[\hat{a}_s,\hat{a}_{s'}]=\delta_{s,s'}$. Replacing discreet sum over modes by a continuous integral $\sum_s\to\frac{V}{(2\pi)^3}\int d\omega_s\tilde{D}(\omega_s)$ with $\tilde{D}(\omega_s)$ being the density of states one can obtain
\begin{align}\label{eq:nVV}
&n(t',\tau)=\mathcal{D}^2(\omega)\langle\mathcal{T} V_R^{\dagger}(t'+\tau)V_L(t')\rangle,
\end{align}
where $\mathcal{D}(\omega)=\frac{1}{2\pi}\tilde{\mathcal{D}}(\omega)$ is a combined density of states evaluated at the central frequency of the detector $\omega$ for smooth enough distribution of modes. The corresponding detected signal (\ref{eq:nDn}) is given by Eq. (\ref{eq:S1g}).

\section{Derivation of the matter correlation expressions for the gated multiple photon counting}\label{app:S2}

 A more elaborate measurement is time-and-frequency resolved photon coincidence counting (PCC) shown in Fig. \ref{fig:diags2}a
\begin{align}\label{eq:S20}
&S^{(2)}(t_1,\omega_1;t_2,\omega_2)\equiv\langle \hat{n}_{t_1,\omega_1}\hat{n}_{t_2,\omega_2}\rangle_T\notag\\
&=\int dt_1' \int d\tau_1 D^{(1)}(t_1\omega_1;t_1',\tau_1)\notag\\
&\times\int dt_2' \int d\tau_2 D^{(2)}(t_2,\omega_2;t_2',\tau_2)\notag\\
&\times\sum_{s,s'}\sum_{r,r'}\langle \hat{E}_{r'}^{\dagger}(t_2'+\tau_2)\hat{E}_{s'}^{\dagger}(t_1'+\tau_1)\hat{E}_s(t_1')\hat{E}_r(t_2')\rangle_T,
\end{align}
where $D^{(j)}$ $j=1,2$ indicates the two detector's spectrograms. The same arguments about temporal and spectral bandwidth discussed about for a single photon detection apply here as well. The corresponding diagram that describes the emission and detection coincidence process is depicted in Fig. \ref{fig:diags2}b. As can be seen from Eq. (\ref{eq:S20}) this constitutes a multidimensional measurement analogous of the $\chi^{(3)}$ nonlinearities detected in rephasing and nonrephasing contributions to the photon echo experiments (see Section \ref{sec:tsj})\cite{Muk95}. Here $\omega_1$ and $\omega_2$ are analogous to Fourier components of the $t$ and $\tau$ delays - $\omega_t$ and $\omega_\tau$, respectively, whereas the time difference between two detectors $t_1-t_2>0$ is analogue of time $t_2$.

Following the same logic regarding the expansion of the signals in the field-matter interactions as in the previous section the second order bare correlation function is given by
\begin{align}\label{eq:Tnn}
S^{(2)}(t_1,\omega_1;t_2,\omega_2)&=\int dt_1' \int d\tau_1 D^{(1)}(t_1\omega_1;t_1',\tau_1)\notag\\
&\times\int dt_2' \int d\tau_2 D^{(2)}(t_2,\omega_2;t_2',\tau_2)\notag\\
&\times\langle \hat{n}(t_1',\tau_1)\hat{n}(t_2',\tau_2')\rangle_T.
\end{align}
The bare PCC rate $\langle\hat{n}(t_1',\tau_1)\hat{n}(t_2',\tau_2')\rangle_T$ can be read off the diagram shown in Fig. \ref{fig:diags2}b. The leading contribution is coming from fourth order expansion over field-matter interactions
\begin{align}
&\langle \mathcal{T}\hat{n}(t_1',\tau_1)\hat{n}(t_2',\tau_2')\rangle_T=\frac{1}{\hbar^4}\int_{-\infty}^{t_1'}dt_1\int_{-\infty}^{t_1'+\tau_1}dt_3\notag\\
&\times\int_{-\infty}^{t_2'}dt_2\int_{-\infty}^{t_2'+\tau_2}dt_4\langle \mathcal{T}V_R^{\dagger}(t_4)V_R^{\dagger}(t_3)V_L(t_1)V_L(t_2)\rangle\notag\\
&\times\sum_{s,s'}\sum_{r,r'}\langle \mathcal{T}E_{r'R}(t_4)E_{s'R}(t_3)E_{r'R}^{\dagger}(t_2'+\tau_2)E_{s'R}^{\dagger}(t_1'+\tau_1)\notag\\
&\times E_{sL}(t_1')E_{rL}(t_2')E_{sL}^{\dagger}(t_1)E_{rL}^{\dagger}(t_2)\rangle_v.
\end{align}
After tracing back the vacuum modes we obtain
\begin{align}\label{eq:nnVVVV}
&\langle\hat{n}(t_1',\tau_1)\hat{n}(t_2',\tau_2')\rangle_T=\mathcal{D}^2(\omega_1)\mathcal{D}^2(\omega_2)\notag\\
&\times\langle\mathcal{T} V_R^{\dagger}(t_2'+\tau_2)V_R^{\dagger}(t_1'+\tau_1)V_L(t_1')V_L(t_2')\rangle.
\end{align}
The corresponding gated coincidence (\ref{eq:Tnn}) is given by Eq. (\ref{eq:S2g}).

\section{Models for gating functions}\label{app:gate}

For Gaussian gates
\begin{align}
F_t(t',t)=e^{-\frac{1}{2}\sigma_T^2(t'-t)^2},\quad F_f(\omega',\omega)=e^{-\frac{(\omega'-\omega)^2}{4\sigma_\omega^2}},
\end{align}
the detector time-domain and Wigner spectrograms are given by
\begin{align}
D(t,\omega,t',\tau)=\frac{\sigma_\omega}{\sqrt{2\pi}}e^{-\frac{1}{2}\sigma_T^2(t'-t)^2-\frac{1}{2}\tilde{\sigma}_\omega^2\tau^2-[\sigma_T^2(t'-t)+i\omega]\tau}
\end{align}
\begin{align}
W_D(t,\omega;t',\omega')=N_De^{-\frac{1}{2}\tilde{\sigma}_T^2(t'-t)^2-\frac{(\omega'-\omega)^2}{2\tilde{\sigma}_\omega^2}-iA(\omega'-\omega)(t'-t)},
\end{align}
where 
\begin{align}
&\tilde{\sigma}_\omega^2=\sigma_T^2+\sigma_\omega^2,\quad \tilde{\sigma}_T^2=\sigma_T^2+\frac{1}{\sigma_\omega^{-2}+\sigma_T^{-2}}, \notag\\
&N_D=\frac{1}{\sigma^T[\sigma_\omega^2+\sigma_T^2]^{1/2}},\quad A=\frac{\sigma_T^2}{\sigma_T^2+\sigma_\omega^2}.
\end{align}
Note that  $\sigma_T$ and $\sigma_\omega$ can be controlled independently and are not subject to uncertainty restrictions, but the actual time and frequency resolution is controlled by $\tilde{\sigma}_T$ and $\tilde{\sigma}_\omega$, respectively, which always satisfy Fourier uncertainty $\tilde{\sigma}_\omega/\tilde{\sigma}_T>1$.
For lorentzian gates
\begin{align}
F_t(t',t)=\theta(t'-t)e^{-\sigma_T(t'-t)},\quad F_f(\omega',\omega)=\frac{i}{\omega'-\omega+i\sigma_\omega},
\end{align}
the detector time-domain and Wigner spectrograms are given by
\begin{align}\label{eq:Dlor}
D(t,\omega,t',\tau)&=\frac{1}{2\sigma_\omega}\theta(t'-t)\theta(t'+\tau-t)e^{-(i\omega+\sigma_T)\tau-2\sigma_T(t'-t)}\notag\\
&\times[\theta(\tau)e^{-\sigma_\omega\tau}+\theta(-\tau)e^{\sigma_\omega\tau}]
\end{align}
\begin{align}\label{eq:WDlor}
&W_D(t,\omega;t',\omega')=-\frac{1}{2\sigma_\omega}\theta(t'-t)e^{-2\sigma_T(t'-t)}\notag\\
&\times\left[\frac{-1}{i(\omega'-\omega)-\sigma_T-\sigma_\omega}+\frac{1-e^{-[i(\omega'-\omega)-\sigma_T+\sigma_\omega](t'-t)}}{i(\omega'-\omega)-\sigma_T+\sigma_\omega}\right].
\end{align}

\section{Calculation of the photon counting signals using two-state-jump model}\label{app:TSJ}

\subsection{Green's functions for the TSJ model}

The total density matrix $\rho$ for TSJ model has 18 components $|\nu\nu's\rangle\rangle$ given by the direct product of nine Liouvills space states $|\nu\nu'\rangle\rangle$ and the two spin states $s=u,d$. The Liouville operator $\hat{\mathcal{L}}$ is diagonal in the vibrational Liouville space variables and is thus given by nine 2$\times$2 blocks in spin space \cite{Sanda1}
\begin{align}
[\hat{\mathcal{L}}]_{\nu\nu's,\nu_1\nu_1's'}=\delta_{\nu\nu_1}\delta_{\nu'\nu_1'}[\hat{L}_S]_{s,s'}+\delta_{\nu\nu1}\delta_{\nu'\nu_1'}\delta_{ss'}[\hat{\mathcal{L}}_S]_{\nu\nu's,\nu\nu's},
\end{align}
where $\hat{L}_S$ describes the two-state-jump kinetics given by
\begin{align}
[\hat{L}_S] =
 \begin{pmatrix}
  -k_d & k_u \\
  k_d & -k_u \\
 \end{pmatrix}.
\end{align}
The up (down) jump rates $k_u$ ($k_d$) are connected by the detailed balance relation $k_u/k_d=\exp\beta(\epsilon_d-\epsilon_u)$ where $\epsilon_d-\epsilon_u$ is the energy difference between the $u$ and $d$ states. 
 The coherent part $\hat{\mathcal{L}}_S=-(i/\hbar)[H_S,...]$ vanishes for the states $|aa\rangle\rangle$, $|cc\rangle\rangle$ blocks $[\hat{\mathcal{L}}_S]_{gg,gg}=[\hat{\mathcal{L}}_S]_{ee.ee}=[\hat{\mathcal{L}}_S]_{ff,ff}=0$. The remaining blocks of $\mathcal{L}_S$ reads
 \begin{widetext}
\begin{align}
[\hat{\mathcal{L}}_S]_{\nu\nu',\nu_1\nu_1'}=\delta_{\nu\nu_1}\delta_{\nu'\nu_1'}
 \begin{pmatrix}
 -i(\epsilon_0^\nu-\epsilon_0^{\nu'})-i(\epsilon_1^\nu-\epsilon_1^{\nu'}) & 0 \\
  0 & -i(\epsilon_0^\nu-\epsilon_0^{\nu'})+i(\epsilon_1^\nu-\epsilon_1^{\nu'}) \\
 \end{pmatrix},
\end{align}
\end{widetext}
where $\epsilon_0^g=0$, $\epsilon_0^e=\Omega_0$, $\epsilon_0^f=2\Omega_0+\Delta_0$, $\epsilon_1^g=0$, $\epsilon_1^e=\Omega_1$, and $\epsilon_1^f=2\Omega_1+\Delta_1$. The two Liouville space Green' functions  $\mathcal{G}(t)=-(i/\hbar)\theta(t)e^{\hat{\mathcal{L}}t}$ - the solution of Eq. (\ref{eq:ME}) are given by \cite{Sanda1}
\begin{align}\label{eq:Gaa}
\mathcal{G}_{gg,gg}(t)&=\mathcal{G}_{ee,ee}(t)=\mathcal{G}_{ff,ff}(t)=(-i/\hbar)\theta(t)\notag\\
&\times\left[\hat{1}+\frac{1-e^{-(k_u+k_d)t}}{k_d+k_u} \begin{pmatrix}
  -k_d& k_u \\
  k_d & -k_u \\
 \end{pmatrix}\right],
 \end{align}
\begin{align}\label{eq:Gac}
\mathcal{G}_{\nu\nu',\nu_1\nu_1'}(t)&=(-i/\hbar)\theta(t)\delta_{\nu\nu'}\delta_{\nu_1\nu_1'}\notag\\
&\times\left[\left(\frac{\eta_2}{\eta_2-\eta_1}\hat{1}-\frac{1}{\eta_2-\eta_1}\hat{\mathcal{L}}_{\nu\nu',\nu\nu'}\right)e^{\eta_1t}\right.\notag\\
&\left.+\left(\frac{\eta_1}{\eta_1-\eta_2}\hat{1}-\frac{1}{\eta_1-\eta_2}\hat{\mathcal{L}}_{\nu\nu',\nu\nu'}\right)e^{\eta_2t}\right],
 \end{align}
where $\hat{1}$ is unit 2$\times$2 matrix and $\eta_{1,2}=-\frac{k_d+k_u}{2}-i(\epsilon_0^\nu-\epsilon_0^{\nu'})\pm\sqrt{\frac{(k_d+k_u)^2}{4}-(\epsilon_1^\nu-\epsilon_1^{\nu'})^2+i(\epsilon_1^\nu-\epsilon_1^{\nu'})(k_d-k_u)}$.


In the low temperature limit $k_BT\ll\epsilon_u-\epsilon_d$ and thus $k_u=0$, Eqs. (\ref{eq:Gaa}) - (\ref{eq:Gac}) are given by the following $2\times 2$ matrices
\begin{align}\label{eq:Gaal}
\mathcal{G}_{gg,gg}(t)&=\mathcal{G}_{ee,ee}(t)=\mathcal{G}_{ff,ff}(t)\notag\\
&=(-i/\hbar)\theta(t)
 \begin{pmatrix}
  e^{-kt}& 0 \\
  1-e^{-kt} & 1 \\
 \end{pmatrix},
 \end{align}
\begin{align}\label{eq:Gacl}
&\mathcal{G}_{eg,eg}(t)=(-i/\hbar)\theta(t)\notag\\
&\times\begin{pmatrix}
  e^{-(k+i\omega_{eg}^{+})t}& 0 \\
  \frac{k}{k+2i\Omega_1}[e^{-i\omega_{eg}^-t}-e^{-(k+i\omega_{eg}^{+})t}] &e^{-i\omega_{eg}^{-}t} \\
 \end{pmatrix},
 \end{align}
\begin{align}\label{eq:Gacl}
&\mathcal{G}_{fe,fe}(t)=(-i/\hbar)\theta(t)\notag\\
&\times\begin{pmatrix}
  e^{-(k+i\omega_{fe}^{+})t}& 0 \\
  \frac{k}{k+2i(\Omega_1+\Delta_1)}[e^{-i\omega_{fe}^-t}-e^{-(k+i\omega_{fe}^{+})t}] &e^{-i\omega_{fe}^{-}t} \\
 \end{pmatrix},
 \end{align}
where $k=k_u$ and $\omega_{eg}^{\pm}=\Omega_0\pm\Omega_1$, $\omega_{fe}^{\pm}=\Omega_0+\Delta_0\pm(\Omega_1+\Delta_1)$.

\subsection{Signals for the TSJ model}

In order to evaluate the signal in  Eq. (\ref{eq:S12}) we first calculate the matter correlation function. Taking into account (\ref{eq:Gaal}) and (\ref{eq:Gacl}) and performing matrix multiplications one obtain for the correlation function in Eq. (\ref{eq:S12})
 \begin{align}\label{eq:S2mcf}
& \langle\langle I|\mathcal{G}_{eg,eg}(\tau)\mathcal{G}_{ee,ee}(t')|\rho_{ee} \rangle\rangle_S=\frac{1}{\hbar^2}\theta(\tau)\theta(t')\rho_{ee}(0)\notag\\
 &\times \left[e^{i\omega_{eg}^-\tau}+\frac{2i\Omega_1}{k-2i\Omega_1}e^{-kt'}(e^{i\omega_{eg}^-\tau}-e^{(-k+i\omega_{eg}^+)\tau})\right].
 \end{align}
Evaluating time integrals in Eq. (\ref{eq:S12}) yields the final result for the signal (\ref{eq:S1tsj}).

The coincidence signal (\ref{eq:S22i}) - (\ref{eq:S22ii}) can be calculated similarly. After calculation of matter correlation function and performing all the time integrals we finally obtain:
 \begin{widetext}
 \begin{align}\label{eq:S23i}
 &S_i^{(2)}(t_1,\omega_1;t_2,\omega_2)=2\mathcal{R}\frac{\mathcal{D}^2(\omega_1)\mathcal{D}^2(\omega_2)}{4\sigma_{f1}\sigma_{f2}}|\mu_{eg}|^2|\mu_{fe}|^2\left[\frac{1}{\Delta_{eg}^-\Delta_{fe}^{-}}\left(\frac{e^{-i\Delta_{fe}^{-}(t_1-t_2)}}{(\Delta_{fe}^{-}-2i\sigma_{T1})(\Delta_{fe}^{-}+2i\sigma_{T2})}-\frac{1}{4\sigma_{T1}\sigma_{T2}}\right)\right.\notag\\
 &\left.+\frac{2i\Omega_1e^{-kt_1}}{(k-2i\Omega_1)\Delta_{fe}^{+}}\left(\frac{1}{\Delta_{eg}^{-}}-\frac{1}{\Delta_{eg}^{+}-ik}\right)\left(\frac{e^{-i\Delta_{fe}^{+}(t_1-t_2)}}{(\Delta_{fe}^+-i(2\sigma_{T1}+k))(\Delta_{fe}^{+}+2i\sigma_{T2})}-\frac{1}{2\sigma_{T2}(2\sigma_{T1}+k)}\right)\right.\notag\\
 &\left.+\frac{2i(\Omega_1+\Delta_1)e^{-kt_2}}{[k-2i(\Omega_1+\Delta_1)]\Delta_{eg}^{-}}\left(\frac{1}{\Delta_{fe}^{-}}\left[\frac{e^{-i\Delta_{fe}^{-}(t_1-t_2)}}{(\Delta_{fe}^{-}-2i\sigma_{T1})(\Delta_{fe}^{-}+i(2\sigma_{T2}+k)}-\frac{1}{2\sigma_{T1}(2\sigma_{T2}+k)}\right]\right.\right.\notag\\
 &\left.\left.-\frac{1}{\Delta_{fe}^{+}-ik}\left[\frac{e^{-(i\Delta_{fe}^{+}+k)(t_1-t_2)}}{(\Delta_{fe}^{+}-i(2\sigma_{T1}+k))(\Delta_{fe}^{+}+2i\sigma_{T2})}-\frac{1}{2\sigma_{T1}(2\sigma_{T2}+k)}\right]\right)\right],
 \end{align}
  \begin{align}\label{eq:S23ii}
 &S_{ii}^{(2)}(t_1,\omega_1;t_2,\omega_2)=2\mathcal{R}\frac{\mathcal{D}^2(\omega_1)\mathcal{D}^2(\omega_2)}{4\sigma_{f1}\sigma_{f2}}|\mu_{eg}|^2|\mu_{fe}|^2\left[\frac{1}{\Delta_{eg}^-\Delta_{fe}^{-*}}\left(\frac{e^{i\Delta_{fe}^{-*}(t_1-t_2)}}{(\Delta_{fe}^{-*}+2i\sigma_{T1})(\Delta_{fe}^{-*}-2i\sigma_{T2})}-\frac{1}{4\sigma_{T1}\sigma_{T2}}\right)\right.\notag\\
 &\left.+\frac{2i\Omega_1e^{-kt_1}}{(k-2i\Omega_1)\Delta_{fe}^{+*}}\left(\frac{1}{\Delta_{eg}^{-}}-\frac{1}{\Delta_{eg}^{+}-ik}\right)\left(\frac{e^{i\Delta_{fe}^{+*}(t_1-t_2)}}{(\Delta_{fe}^{+*}+i(2\sigma_{T1}+k))(\Delta_{fe}^{+*}-2i\sigma_{T2})}-\frac{1}{2\sigma_{T2}(2\sigma_{T1}+k)}\right)\right.\notag\\
 &\left.+\frac{2i(\Omega_1+\Delta_1)e^{-kt_2}}{[k+2i(\Omega_1+\Delta_1)]\Delta_{eg}^{-}}\left(\frac{1}{\Delta_{fe}^{-*}}\left[\frac{e^{i\Delta_{fe}^{-*}(t_1-t_2)}}{(\Delta_{fe}^{-*}+2i\sigma_{T1})(\Delta_{fe}^{-*}-i(2\sigma_{T2}+k)}-\frac{1}{2\sigma_{T1}(2\sigma_{T2}+k)}\right]\right.\right.\notag\\
 &\left.\left.-\frac{1}{\Delta_{fe}^{+*}+ik}\left[\frac{e^{(i\Delta_{fe}^{+*}-k)(t_1-t_2)}}{(\Delta_{fe}^{+*}+i(2\sigma_{T1}+k))(\Delta_{fe}^{+*}-2i\sigma_{T2})}-\frac{1}{2\sigma_{T1}(2\sigma_{T2}+k)}\right]\right)\right],
 \end{align}
\end{widetext}
where $\Delta_{fe}^{\pm}=\omega_2-\omega_{fe}^{\pm}-i(\sigma_{T2}+\sigma_{\omega2})$.


%

\end{document}